\documentclass[envcountsame]{llncs}

\usepackage{pscproc2}

\usepackage{color}
\usepackage{epic,eepic}

\usepackage{epsfig}
\usepackage{amsfonts}
\usepackage{latexsym}
\usepackage{amssymb}
\usepackage{amsmath}

\usepackage{float}
\usepackage{xspace}

\usepackage{url}

\newcommand{\uassign}{\leftarrow}
\newcommand{\uif}{{\bf if}\xspace}
\newcommand{\uthen}{{\bf then}\xspace}
\newcommand{\uelse}{{\bf else}\xspace}

\newcommand{\ufor}{{\bf for}\xspace}
\newcommand{\uto}{{\bf to}\xspace}

\newcommand{\udo}{{\bf do}\xspace}

\newcounter{lineno}
\newcommand{\uln}{\stepcounter{lineno}\>\hspace{-5mm}\thelineno}
\newcommand{\utab}{\qquad}
\newcommand{\utrue}{\mbox{\textbf{true}}\xspace}
\newcommand{\ufalse}{\mbox{\textbf{false}}\xspace}
\newenvironment{code}{%
\scriptsize
\setcounter{lineno}{0}%
\begin{tabbing}
\utab\=\utab\=\utab\=\utab\=\utab\=\utab\=\utab\=%
\utab\=\utab\=\utab\=\utab\=\utab\=\utab\= \kill
}
{
\end{tabbing}%
\vspace{-4mm}
}

\floatstyle{ruled}
\newfloat{algorithm}{tbp}{flt}
\floatname{algorithm}{\small\bf Alg.}

\newcommand{\exclude}[1]{}

\begin{document}

%%%%%%%%%%%%%%%%%%%%%%%%%%%%%%%%%%%%%%%%%%%%%%%%%%%%%%%%%%%%%%%%%%%%%%%%%%%%%%
\title{Tight and simple Web graph compression}
%%%%%%%%%%%%%%%%%%%%%%%%%%%%%%%%%%%%%%%%%%%%%%%%%%%%%%%%%%%%%%%%%%%%%%%%%%%%%%

\author{
Szymon Grabowski
and
Wojciech Bieniecki
}

\institute{
Computer Engineering Department, Technical University of {\L}\'od\'z,\\
Al.\ Politechniki 11, 90--924 {\L}\'od\'z, Poland \\
\email{\{sgrabow,wbieniec\}@kis.p.lodz.pl}
}

\maketitle

\begin{abstract}
Analysing Web graphs has applications in determining page ranks, 
fighting Web spam, detecting communities and mirror sites, and more.
This study is however hampered by the necessity of storing a major part 
of huge graphs in the external memory, which prevents efficient random 
access to edge (hyperlink) lists.
A number of algorithms involving compression techniques have thus been 
presented, to represent Web graphs succinctly but also providing random 
access.  Those techniques are usually based on differential encodings of 
the adjacency lists, finding repeating nodes or node regions in the 
successive lists, more general grammar-based transformations or 
2-dimensional representations of the binary matrix of the graph.
In this paper we present 
two Web graph compression algorithms. The first can be seen 
as engineering of the Boldi and Vigna (2004) method.
We extend the notion of similarity between link lists, and use a more 
compact encoding of residuals.  The algorithm works on blocks of varying 
size (in the number of input lines) and sacrifices access time for better 
compression ratio, achieving more succinct graph representation than 
other algorithms reported in the literature.
The second algorithm works on blocks of the same size, in the number of 
input lines, and its key mechanism is merging the block into a single ordered 
list. This method achieves much more attractive space-time tradeoffs.
\end{abstract}


%%%%%%%%%%%%%%%%%%%%%%%%%%%%%%%%%%%%%%%%%%%%%%%%%%%%%%%%%%%%%%%%%%%%%%%%%%%%%%
\section{Introduction}
%%%%%%%%%%%%%%%%%%%%%%%%%%%%%%%%%%%%%%%%%%%%%%%%%%%%%%%%%%%%%%%%%%%%%%%%%%%%%%

Development of succinct data structures is one of the most active research areas 
in algorithmics in the last years.  A succinct data structure shares the interface 
with its classic (non-succinct) counterpart, but is represented in much smaller 
space, via data compression.
%% techniques.
Successful examples along these lines include text indexes \cite{NMacmcs06}, 
dictionaries, trees \cite{MR97,GearyRRR04} and graphs \cite{MR97}.
%% unlabeled graphs \cite{MR97}, labeled graphs \cite{BCAHM07}
Queries to succinct data structures are usually slower (in practice, although 
not always in complexity terms) than using non-compressed structures, hence 
the main motivation in using them is to allow to deal with huge datasets in the 
main memory.
For example, indexed exact pattern matching in DNA would be limited to 
sequences shorter than 1 billion nucleotides on a commodity PC with 4~GB of 
main memory, if the indexing structure were the classic suffix array (SA), 
and even less than half of it, if SA were replaced with a suffix tree.  
On the other hand, switching to some compressed full-text index (see 
\cite{NMacmcs06} for a survey) shifts the limit to over 10 billion nucleotides, 
which is more than enough to handle the whole human genome.

Another huge object of significant interest seems to be the Web graph.  
This is a directed unlabeled graph of connections between webpages (i.e., documents), 
where the nodes are individual HTML documents and the edges from a given node 
are the outgoing links to other nodes.  We assume that the order of hyperlinks 
in a document is irrelevant.  
Web graph analyses can be used to rank pages, fight Web spam, detect communities 
and mirror sites, etc. 

As of early Sept. 2011, it is estimated that Google's index has about 44 billion 
webpages\footnote{\url{http://www.worldwidewebsize.com/}}.
Assuming 20 outgoing links per node, 5-byte links (4-byte indexes to other pages
are simply too small) and pointers to each adjacency list, we would need more 
than 4.4\,TB of memory, ways beyond the capacities of the current RAM memories.
We believe that, confronted with the given figures, the reader is now convinced 
about the necessity of compression techniques for Web graph representation.

Preliminary versions of this manuscript were published in~\cite{GB10} 
and~\cite{GB11}.
%% Prague Stringology Conference 2010 and ICMMI'11

%%%%%%%%%%%%%%%%%%%%%%%%%%%%%%%%%%%%%%%%%%%%%%%%%%%%%%%%%%%%%%%%%%%%%%%%%%%%%%
\section{Related work}
\label{sec:relatedwork}
%%%%%%%%%%%%%%%%%%%%%%%%%%%%%%%%%%%%%%%%%%%%%%%%%%%%%%%%%%%%%%%%%%%%%%%%%%%%%%

We assume that a directed graph $G = (V, E)$ is a set of $n = |V|$ vertices and 
$m = |E|$ edges.
The earliest works on graph compression were theoretical, and they usually dealt 
with specific graph classes.  For example, it is known that planar graphs can be 
compressed into $O(n)$ bits \cite{Turan84,DBLP:journals/siamcomp/HeKL00}.  
For dense enough graphs, it is impossible to reach $o(m \log n)$ bits of space, 
i.e., go below the space complexity of the trivial adjacency list representation.
Since the seminal Jacobson's thesis \cite{Jac89} on succinct data 
structures, there appear papers taking into account not only the space occupied 
by a graph, but also access times.

There are several works dedicated to Web graph compression.  Bharat et al. 
\cite{DBLP:journals/cn/BharatBHKV98}
suggested to order documents according to their URL's, to exploit the simple 
observation that most outgoing links actually point to another document within 
the same Web site.  Their Connectivity Server provided linkage information 
for all pages indexed by the AltaVista search engine at that time.
The links are merely represented by the node numbers (integers) using the URL 
lexicographical order.  We noted that we assume the order of hyperlinks in a 
document irrelevant (like most works on Web graph compression do), hence 
the link lists can be sorted, in ascending order.
As the successive numbers tend to be close, differential encoding may be applied 
efficiently.

Randall et al. \cite{randall01link} also use this technique (stating that for 
their data 80\% of all links are local), but they also note that commonly many 
pages within the same site share large parts of their adjacency lists.  
To exploit this phenomenon, a given list may be encoded with a reference to another 
list from its neighborhood (located earlier), plus a set of additions and deletions 
to/from the referenced list.
Their encoding, in the most compact variant, encodes an outgoing link in 5.55 bits 
on average, a result reported over a Web crawl consisting of 61 million URL's and 
1 billion links.

One of the most efficient compression schemes for Web graph was presented by
Boldi and Vigna \cite{DBLP:conf/www/BoldiV04} in 2003. Their method is likely to 
achieve around 3 bits per edge, or less, at link access time below 1\,ms at their 
2.4\,GHz Pentium4 machine.  Of course, the compression ratios vary from dataset 
to dataset. We are going to describe the Boldi and Vigna algorithm in detail in 
the next section as this is the main inspiration for our solution.

Claude and Navarro \cite{CNtr08,CNtweb10} took a totally different approach of grammar-based 
compression. In particular, they focus on Re-Pair \cite{OffDict2000} and LZ78 
compression schemes, getting close, and sometimes even below, the compression 
ratios of Boldi and Vigna, while achieving much faster access times.
To mitigate one of the main disadvantages of Re-Pair, high memory requirements, 
they developed an approximate variant of this algorithm.

When compression is at a premium, one may acknowledge the work of Asano et al. 
\cite{DBLP:conf/cocoon/AsanoMN08}
in which they present a scheme creating a compressed graph structure smaller 
by about 20--35\% than the BV scheme with extreme parameters (best compression but 
also impractically slow). The Asano et al. scheme perceives the Web graph 
as a binary matrix (1s stand for edges) and detects 2-dimensional redundancies 
in it, via finding six types of blocks in the matrix: horizontal, vertical, diagonal, 
L-shaped, rectangular and singleton blocks. The algorithm compresses the data 
of intra-hosts separately for each host, and the boundaries between hosts must be 
taken from a separate source (usually, the list of all URL's in the graph), hence 
it cannot be justly compared to other algorithms mentioned here. 
Worse, retrieval times per adjacency list are much longer than for other schemes: 
on the order of a few milliseconds (and even over 28\,ms for one of three tested 
datasets) on their Core2 Duo E6600 (2.40\,GHz) machine running Java code.
We note that 28\,ms is at least twice more than the access time of modern hard disks, 
hence working with a na\"ive (uncompressed) external representation would be 
faster for that dataset (on the other hand, excessive disk use from very frequent
random accesses to the graph can result in a premature disk failure).
It seems that the retrieval times can be reduced (and made more stable across 
datasets) if the boundaries between hosts in the graph are set artificially, 
in more or less regular distances, but then also the compression ratio is likely 
to drop. 

Also excellent compression results were achieved by Buehrer and Chellapilla
\cite{DBLP:conf/wsdm/BuehrerC08}, who used grammar-based compression. 
Namely, they replace groups of nodes appearing in several adjacency lists with a 
single ``virtual node'' and iterate this procedure; no access times were reported 
in that work, but according to findings in \cite{DBLP:conf/birthday/ClaudeN10} 
they should be rather competitive and at least much shorter than 
of the algorithm from \cite{DBLP:conf/cocoon/AsanoMN08}, with compression ratio 
worse only by a few percent.

Apostolico and Drovandi~\cite{AD09} proposed an alternative Web graph 
ordering, reflecting their BFS traversal (starting from a random node) rather 
than traditional URL-based order.
They obtain quite impressive compressed graph structures, often by 20--30\% smaller 
than those from BV at comparable access speeds.
Interestingly, the BFS ordering allows to handle the link existential query 
(testing if page $i$ has a link to page $j$) almost twice faster than returning 
the whole neighbor list.
Still, we note that using non-lexicographical ordering is harmful 
for compact storing of the webpage URLs themselves (a problem accompanying pure 
graph structure compression in most practical applications).
Note also that reordering the graph is the approach followed in more recent 
works from the Boldi and Vigna team~\cite{BSV09,BoldiRSV11}.

Anh and Moffat \cite{DBLP:conf/dcc/AnhM10} devised a scheme which seems to use 
grammar-based compression in a local manner. They work in groups of $h$ consecutive 
lists and perform some operations to reduce their size (e.g., a sort of 2-dimensional 
RLE if a run of successive integers appears on all the $h$ lists).  What remains 
in the group is then encoded statistically.  Their results are very promising: 
graph representations by about 15--30\% (or even more in some variant) smaller 
than the BV algorithm with practical parameter choice (in particular, 
Anh and Moffat achieve 3.81\,bpe and 3.55\,bpe for the graph EU) and report 
comparable decoding speed.
Details of the algorithm cannot however be deduced 
from their 1-page conference poster.

Recent works focus on graph compression with support for bidirectional navigation.
To this end, Brisaboa et al. \cite{BLNspire09.1} proposed the {\em $k^2$-tree}, 
a spatial data structure, 
related to the well-known quadtree, which performs a binary partition of the graph 
matrix and labels empty areas with 0s and non-empty areas with 1s. The non-empty 
areas are recursively split and labeled, until reaching the leaves (single nodes).
An important component in their scheme is an auxiliary structure to compute {\em rank}
queries \cite{Jac89} efficiently, to navigate between tree levels.
It is easy to notice that this elegant data structure supports handling both forward  
and reverse neighbors, which implies from its symmetry. 
Ladra~\cite{L11} proposed a more efficient encoding of leaves (which 
are boxes of sizes e.g. $8 \times 8$ rather than single bits)
in this scheme, making use of a common vocabulary for the different 
leaf submatrices and directly addressable codes.
Very recently, on the base of the mentioned encoding, Claude and Ladra~\cite{CL11} 
achieved even better results, and the key idea was to divide the original square matrix 
into subdomains, cutting out several non-overlapping squares (subgraphs) 
along the diagonal of the binary matrix; each generated subgraph is stored independently.
Experiments show that even the original work uses significantly less space 
(3.3--5.3 bits per link) than the Boldi and Vigna 
scheme applied for both direct and transposed graph, at the average neighbor retrieval 
times of 2--15 microseconds (Pentium4 3.0\,GHz).
The Claude and Ladra variant reduces the space to about 3--4 bits per link 
and the retrieval time is improved to about 1 microsecond or less (Intel Xeon 2.0\,GHz).

In other recent work, Claude and Navarro \cite{DBLP:conf/birthday/ClaudeN10} showed 
how Re-Pair can be used to compress the graph binary relation efficiently, enabling 
also to extract the reverse neighbors of any node. These ideas let them achieve 
a number of Pareto-optimal space-time tradeoffs, usually competitive to those from 
the (original variant of the) $k^2$-tree.

Finally, we have to mention the Hern{\'a}ndez and Navarro work~\cite{HNsnakdd11}, 
where they combine their previous techniques, $k^2$-tree~\cite{BLNspire09.1} 
and Re-Pair for compressing the graph binary relation~\cite{DBLP:conf/birthday/ClaudeN10} 
with edge reducing~\cite{DBLP:conf/wsdm/BuehrerC08}, obtaining interesting trade-offs.
In particular, if some of the access time can be sacrified, the space they achieved is 
the smallest known among the solutions supporting bidirectional queries.

%%%%%%%%%%%%%%%%%%%%%%%%%%%%%%%%%%%%%%%%%%%%%%%%%%%%%%%%%%%%%%%%%%%%%%%%%%%%%%
\section{The Boldi and Vigna scheme}
\label{sec:bv}
%%%%%%%%%%%%%%%%%%%%%%%%%%%%%%%%%%%%%%%%%%%%%%%%%%%%%%%%%%%%%%%%%%%%%%%%%%%%%%

Based on WebGraph datasets (\url{http://webgraph.dsi.unimi.it/}), Boldi and Vigna noticed 
that similarity is strongly concentrated; typically, either two adjacency (edge) lists 
have nothing or little in common, or they share large subsequences of edges.  
To exploit this redudancy, one bit per entry on the referenced list could be used, 
to denote which of its integers are copied to the current list, and which are not.  
Those bit-vectors are dubbed {\em copy lists}.  Still, Boldi and Vigna go further, 
noticing that copy lists tend to contain runs of 0s and 1s, thus they compress them 
using a sort of run-length encoding.  They assume the first run consists of 1s 
(if the copy list actually starts with 0s, the length of the first run is simply zero), 
and then it allows to represent a copy list as only a sequence of run lengths, 
encoded e.g. with Elias coding.

The integers on the current list which didn't occur on the referenced list must 
be stored too, and how to encode them is another novelty of the described 
algorithm.  They detect intervals of consecutive (i.e., differing by 1) integers
and encode them as pairs of the left boundary and the interval length; 
the left boundary of the next interval on a given list will be encoded as the 
difference to the right boundary of the previous interval minus two (this is 
because between the end of one interval and the beginning of another there 
must be at least one integer).
The numbers which do not fall into any interval are called {\em residuals} 
and are also stored, encoded in a differential manner.

Finally, the algorithm allows to select as the reference list one of several 
previous lines; the size of the {\em window} is one of the parameters of the 
algorithm posing a tradeoff between compression ratio and compression/decompression 
time and space.
Another parameter affecting the results is the maximum reference count, 
which is the maximum allowed length of a chain of lists such that one cannot be 
decoded without extracting its predecessor in the chain.

%%%%%%%%%%%%%%%%%%%%%%%%%%%%%%%%%%%%%%%%%%%%%%%%%%%%%%%%%%%%%%%%%%%%%%%%%%%%%%
\section{Our algorithms}
\label{sec:our}
%%%%%%%%%%%%%%%%%%%%%%%%%%%%%%%%%%%%%%%%%%%%%%%%%%%%%%%%%%%%%%%%%%%%%%%%%%%%%%

We present two approaches to Web graph compression working locally, in small 
blocks; the first one usually reaches slightly higher compression ratios 
but the second is more practical, as being much faster.

%%%%%%%%%%%%%%%%%%%%%%%%%%%%%%%%%%%%%%%%%%%%%%%%%%%%%%%%%%%%%%%%%%%%%%%%%%%%%%
\subsection{An algorithm based on similarity of successive lists}
\label{sec:fa}
%%%%%%%%%%%%%%%%%%%%%%%%%%%%%%%%%%%%%%%%%%%%%%%%%%%%%%%%%%%%%%%%%%%%%%%%%%%%%%

Our first algorithm (Alg.~\ref{alg:GraphC1}, SSL stands for ``similarity of successive 
lists'') works in blocks consisting of multiple adjacency lists. The blocks in their 
compact form are approximately equal, which means that the number of adjacency lists 
per block varies; for example, in graph areas with dominating short lists the number 
of lists per block is greater than elsewhere.

We work in two phases: preprocessing and final compression, using a general-purpose 
compression algorithm. The algorithm processes the adjacency lines one-by-one and 
splits their data into two streams.

\begin{algorithm}[!t]
\scriptsize
\footnotesize
\begin{code}
\uln \> $firstLine \uassign \utrue$ \\
\uln \> $prev \uassign$ [\ ] \\
\uln \> $outB \uassign$ [\ ] \\
\uln \> $outF \uassign$ [\ ] \\
\uln \>	\ufor $line \in G$ \udo \\
\uln \>	\> $residuals \uassign line$ \\
\uln \> \> \uif $firstLine = \ufalse$ \uthen \\
\uln \>	\> \> $f[1 \ldots |prev|] \uassign [1, 1, \ldots, 1]$ \\
\uln \>	\> \> \ufor $i \uassign 1$ \uto $|prev|$ \udo \\
\uln \>	\> \> \> \uif $prev[i] \in line$ \uthen $f[i] \uassign 0$ \\
\uln \>	\> \> \> \uelse~\uif $prev[i]+1 \in line$ \uthen $f[i] \uassign 2$ \\
\uln \>	\> \> \> \uelse~\uif $prev[i]+2 \in line$ \uthen $f[i] \uassign 3$ \\
\uln \>	\> \> append($outF$, $f$) \\
\uln \>	\> \> \ufor $i \uassign 1$ \uto $|prev|$ \udo \\
\uln \>	\> \> \> \uif $f[i] \neq 1$ \uthen \\ 
\uln \>	\> \> \> \> remove($residuals$, $prev[i]$) \\
\uln \>	\> $residuals' \uassign$ RLE(diffEncode($residuals$)) + [0] \\
\uln \>	\> append($outB$, byteEncode($residuals'$)) \\
\uln \>	\> $prev \uassign line$ \\
\uln \>	\> $firstLine \uassign \ufalse$ \\
\uln \> \> \uif $|outB| \geq BSIZE$ \uthen \\
\uln \> \> \> compress($outB$) \\
\uln \> \> \> compress($outF$) \\
\uln \> \> \> $outB \uassign$ [\ ] \\
\uln \> \> \> $outF \uassign$ [\ ] \\
\uln \> \> \> $firstLine \uassign \utrue$ \\
\end{code}
\caption{GraphCompressSSL($G, BSIZE$).}
\label{alg:GraphC1}
\end{algorithm}

One stream holds copy lists, in an extended sense compared to the 
Boldi and Vigna solution.  Our copy lists are no longer binary but consist of 
four different flag symbols: 0 denotes an exact match (i.e., value $j$ from the 
reference list occurs somewhere on the current list), 2 means that the current list 
contains integer $j+1$, 3 means that the current list contains integer $j+2$, 
if the corresponding integer from the reference list is $j$.
Finally, the bits 1 correspond to the items from the reference list which have not 
been earlier labeled with 0, 2 or 3.

Of course, several events may happen for a single 
element, e.g., the integer 34 from the reference list triggers three events 
if the current list contains 34, 35 and 36.  In such case, the flag with the 
smallest value is chosen (i.e., 0 in our example).

Moreover, we make things even simpler than in the Boldi--Vigna scheme 
and our reference list is always the previous adjacency list.

The other stream stores residuals, i.e., the values which cannot be decoded 
with flags 0, 2 or 3 on the copy lists. First differential encoding is applied 
and then an RLE compressor for differences 1 only (with minimum run length set 
experimentally to 5) is run. The resulting sequence is terminated with a unique 
value (0) and then encoded using a byte code. 

For this last step, we consider two variants. One is similar to {\em two-byte 
dense code} \cite{DBLP:conf/iwoca/ProchazkaH09} 
in spending one bit flag in the first codeword byte to tell the length of the current 
codeword.  Namely, we choose between 1 and $b$ bytes for encoding each number, 
where $b$ is the minimum integer such that $8b - 1$ bits are enough to encode any 
node value in a given graph. In practice it means that $b = 3$ for EU and $b = 4$ for 
the remaining available datasets.

The second coding variant can be classified as a prelude 
code \cite{DBLP:conf/spire/CulpepperM05} in which two bits in the first codeword byte 
tell the length of the current codeword; originally the lengths are 1, 2, 3 and 4 
but we take 1, 2 and $b$ 
such that $8b - 2$ bits are enough to encode the 
largest value in the given graph (i.e., $b$ could be 5 or 6 for really huge graphs).

Once the residual buffer reaches at least BSIZE bytes, it is time to end the current 
block and start a new one.  Both residual and flag buffers and then (independently) 
compressed (we used the well-known Deflate algorithm for this purpose) and flushed. 

The code at Alg.~\ref{alg:GraphC1} is slightly simplified; we omitted technical details 
serving for finding the list boundaries in all cases (e.g., empty lines).

%%%%%%%%%%%%%%%%%%%%%%%%%%%%%%%%%%%%%%%%%%%%%%%%%%%%%%%%%%%%%%%%%%%%%%%%%%%%%%
\subsection{An algorithm based on list merging}
\label{sec:sa}
%%%%%%%%%%%%%%%%%%%%%%%%%%%%%%%%%%%%%%%%%%%%%%%%%%%%%%%%%%%%%%%%%%%%%%%%%%%%%%

Our second algorithm (Alg.~\ref{alg:GraphC2}, LM stands for ``list merging'') 
works in blocks having the same number of lists, $h$ (at least in this aspect 
our algorithm resembles the one from \cite{DBLP:conf/dcc/AnhM10}).

Given the block of $h$ lists, the procedure converts it into two streams: one stores 
one long list consisting of all integers on the $h$ input lists, without duplicates, 
and the other stores flags necessary to reconstruct the original lists.
In other words, the algorithm performs a reversible merge of all the lists 
in the block.

The long list is compacted in a manner similar to the previous algorithm:
the list is differentially encoded, zero-terminated and submitted to a byte coder 
(the variant with 1, 2 and $b$ bytes per codeword was only tried). Note we gave up 
the RLE phase here.

The flags describe to which input lists a given integer on the output list belongs; 
the number of bits per each item on the output list is $h$, and in practical terms 
we assume $h$ being a multiple of 8 (and even additionally a power of 2, in the 
experiments to follow). 
The flag sequence does not need any terminator since its length is defined by the 
length of the long list, which is located earlier in the output stream.
For example, if the length of the long list is 91 and $h=32$, the corresponding 
flag sequence has 364 bytes.

Now, we consider two variations for encoding the flag sequence:
either they are kept raw (the variant is latter denoted as {\em LM-bitmap}), 
or differences (gaps) between the successive 1s in the flag sequence are written 
on individual bytes (the variant is latter denoted as {\em LM-diff}).
We note that each run of $h$ bits corresponding to flags for a single value on the 
output list must contain at least one set bit, hence the maximum gap 
between any two 1s in the resulting sequence is $2h - 1$, hence for $h \leq 128$ 
each value can be stored on a byte 
(a preliminary experiment with $h = 256$ and using a byte code for gap encoding 
was rather unsucessful).
Alg.~\ref{alg:GraphC2} presents the LM-bitmap variant.

Those two sequences, the compacted long list and the flag sequence (either raw, 
or gap-encoded), are then concatenated and compressed with the Deflate algorithm.

\begin{algorithm}[!t]
\scriptsize
\footnotesize
\begin{code}
\uln \> $outF \uassign$ [\ ] \\
\uln \> $i \uassign$ 1 \\
\uln \>	\ufor $line_i, line_{i+1}, \ldots, line_{i+h-1} \in G$ \udo \\
\uln \>	\> $tempLine_1 \uassign line_i \cup line_{i+1} \cup \ldots \cup line_{i+h-1}$ \\
\uln \> \> $tempLine_2 \uassign$ removeDuplicates($tempLine_1$) \\
\uln \>	\> $longLine \uassign$ sort($tempLine_2$) \\
\uln \>	\> $items \uassign$ diffEncode($longLine$) + [0] \\
\uln \>	\> $outB \uassign$ byteEncode($items$) \\
\uln \>	\> \ufor $j \uassign 1$ \uto $|longLine|$ \udo \\
\uln \>	\> \> $f[1 \ldots |longLine|] \uassign [0, 0, \ldots, 0]$ \\
\uln \>	\> \> \ufor $k \uassign 1$ \uto $h$ \udo \\
\uln \>	\> \> \> \uif $longLine[j] \in line_{i+k-1}$ \uthen $f[k] \uassign 1$ \\
\uln \>	\> \> append($outF$, bitPack($f$)) \\
\uln \> \> compress(concat($outB$, $outF$)) \\
\uln \> \> $outF \uassign$ [\ ] \\
\uln \> \> $i \uassign i+h$ \\
\end{code}
\caption{GraphCompressLM($G, h$).}
\label{alg:GraphC2}
\end{algorithm}

One can see that the key parameter here is the block size, $h$. 
Using a larger $h$ lets exploit a wider range of similar lists but also has two 
drawbacks. 
The flag sequence gets more and more sparse (for example, for $h=64$ 
and the EU-2005 crawl, as much as about 68\% of its list indicators have only one 
set bit out of 64!), and the Deflate compressor is becoming relatively inefficient 
on those data; a drawback more important in the LM-bitmap variant.
Worse, decoding larger blocks takes longer time.

%%%%%%%%%%%%%%%%%%%%%%%%%%%%%%%%%%%%%%%%%%%%%%%%%%%%%%%%%%%%%%%%%%%%%%%%%%%%%%
\section{Experimental results}
%%%%%%%%%%%%%%%%%%%%%%%%%%%%%%%%%%%%%%%%%%%%%%%%%%%%%%%%%%%%%%%%%%%%%%%%%%%%%%

The experiments with the SSL algorithm comprise only the datasets 
EU-2005 and Indochina-2004, while the more practical LM variants 
are tested also on the UK-2002 and Arabic-2005 crawls; 
all the datasets are 
downloaded from the WebGraph project (\url{http://webgraph.dsi.unimi.it/}),
using both direct and transposed graphs.
Note that we use the natural order versions of them, as 
using reordered variants (also available from the WebGraph project) 
may be more efficient but then the compression of the corresponding URL data 
deteriorates.

The main characteristics of those datasets are presented in Table~\ref{table:datasets}.

\begin{table}
\centering
\begin{tabular}{lrrrrrrrr}
\hline
Dataset & \multicolumn{2}{c}{EU-2005} &  \multicolumn{2}{c}{Indochina-2004} & \multicolumn{2}{c}{UK-2002} &  \multicolumn{2}{c}{Arabic-2005} \\
\cline{2-9}
        &  direct & transposed & direct & transposed  
        &  direct & transposed & direct & transposed \\
\hline
Nodes & 862664   & & 7414866   & & 18520486  & & 22744080  & \\
Edges & 19235140 & & 194109311 & & 298113762 & & 639999458 & \\
Edges / nodes & 22.30 & & 26.18 & & 16.10 & & 28.14 & \\
\% of empty lists & 8.309 & 0.000 & 17.655 & 0.004 & 14.908 & 0.637 & 14.514 &  0.002 \\
Longest list length & 6985 & 68922 & 6985 & 256425 & 2450 & 194942 & 9905 & 575618 \\
\hline
\end{tabular}
\vspace{4mm}
\caption{Selected characteristics of the datasets used in the experiments.}
\label{table:datasets}
\end{table}

The main experiments (Sect.~\ref{sec:craat})
were run on a machine equipped with an Intel Core 2 Quad Q9450 CPU, 
8 GB of RAM, running Microsoft Windows XP (64-bit).
Our algorithms were implemented in Java and run on the 64-bit JVM 
(JRE 6 used in the first series of tests, involving SSL, 
and JRE 7 in the latter tests, with the LM variants). 
A single CPU core was used by all implementations.
As seemingly accepted in most reported works, 
we measure access time per edge, extracting many (100,000 in our case) 
randomly selected adjacency lists and summing those times, and dividing the 
total time by the number of edges on the required lists.
The space is measured in bits per edge (bpe), dividing the
total space of the structure (including entry points to blocks)
by the total number of edges.

Throughout this section by 1\,KB we mean 1000 bytes.

%%%%%%%%%%%%%%%%%%%%%%%%%%%%%%%%%%%%%%%%%%%%%%%%%%%%%%%%%%%%%%%%%%%%%%%%%%%%%%
\subsection{Compression ratios and access times}
\label{sec:craat}
%%%%%%%%%%%%%%%%%%%%%%%%%%%%%%%%%%%%%%%%%%%%%%%%%%%%%%%%%%%%%%%%%%%%%%%%%%%%%%

Our first algorithm, SSL, has three parameters: the number of flags used (either 2 or 4, 
where 2 flags mimic the Boldi--Vigna scheme and 4 correspond to Alg.~\ref{alg:GraphC1}),
the byte encoding scheme (either using 2 or 3 codeword lengths), and the residual
block size threshold BSIZE. As for the last parameter, we initially set it to 8192, 
which means that the residual block gets closed and is submitted to the Deflate 
compression once it reaches at least 8192 bytes. Experiments with the block size 
are presented in the next subsection. 
The remaining parameters constitute four variants:
\begin{description}
  \item[2a] Two flags and two codeword lengths are used.
  \item[2b] Two flags and three codeword lengths are used.
  \item[4a] Four flags and two codeword lengths are used.
  \item[4b] Four flags and three codeword lengths are used.
\end{description}

\begin{table}
\centering
\begin{tabular}{lrrrr}
\hline
Dataset & \multicolumn{2}{c}{EU-2005} &  \multicolumn{2}{c}{Indochina-2004}  \\
\cline{2-5}
        &  direct & transposed~~~&~~~direct & transposed  \\
\hline
2a & 2.286 & 2.345 & 1.101 & 1.087 \\
2b & 2.199 & 2.290 & 1.062 & 1.065 \\
4a & 1.735 & 1.809 & 0.936 & 0.903 \\
4b & 1.696 & 1.782 & 0.909 & 0.890 \\
\hline
\end{tabular}
\vspace{4mm}
\caption{The algorithm based on similarity of successive lists, compression ratios in bits per edge.}
\label{table:ratios}
\end{table}

As expected, the compression ratios improve with using more flags and more dense 
byte codes (Table~\ref{table:ratios}).
Tables~\ref{table:eu} and \ref{table:indochina} present the compression 
and access time results for the two extreme variants: 2a and 4b.
Here we see that using more aggressive preprocessing is unfortunately slower 
(partly because of increased amount of flag data per block)
and the difference in speed between variants 2a and 4b is close to 50\%. 
Translating the times per edge into times per neighbor list, we need from 
410\,$\mu$s to 550\,$\mu$s for 2a and from 620\,$\mu$s to 760\,$\mu$s for 4b. 
This is about 10 times less than the access time of 10K or 15K RPM hard disks.

Our second algorithm, LM, has one parameter, $h$, the number of lines (lists) per block.
We conducted experiments for $h = 16$, 32, 64, the results are presented in the 
last three rows of Tables~\ref{table:eu} and \ref{table:indochina}, respectively.
For this comparison, only the LM-bitmap variant is used.
We see that even LM64 cannot reach the compression of our 4b variant, but its list 
extraction is faster 14--27 times. The fastest of the variants presented here, LM16, 
is 1.3 and 2.0 slower than BV (7,3), respectively, with much better compression 
(we checked also LM8, only on EU-2005: the results are 3.814\,bpe and 0.20\,$\mu$s 
per edge).

\begin{table}
\centering
\begin{tabular}{lrrrr}
\hline
        & \multicolumn{2}{c}{direct graph} & \multicolumn{2}{c}{transposed graph} \\
\cline{2-5}
        &  bpe~~~&~~~time [$\mu$s]~~~&~~~bpe~~~&~~~time [$\mu$s] \\
\hline
BV (7,3) & 5.169 & 0.24 & -- & -- \\
2a & 2.286 & 18.59 & 2.345 & 18.88 \\
4b & 1.696 & 28.93 & 1.782 & 27.83 \\
LM16 & 2.963 & 0.31 & 2.576 & 0.82 \\
LM32 & 2.373 & 0.55 & 2.233 & 1.05 \\
LM64 & 2.008 & 1.05 & 2.016 & 2.01 \\
\hline
\end{tabular}
\vspace{4mm}
\caption{EU-2005 dataset. Compression ratios (bpe) and access times per edge. 
``LM$x$'' stands for LM-bitmap with $h = x$.
To the results of BV (7,3) the amount of 0.510\,bpe should be added, corresponding 
to extra data required to access the graph in random order.}
\label{table:eu}
\end{table}

\begin{table}
\centering
\begin{tabular}{lrrrr}
\hline
        & \multicolumn{2}{c}{direct graph} & \multicolumn{2}{c}{transposed graph} \\
\cline{2-5}
     ~~~&~~~bpe~~~&~~~time [$\mu$s]~~~&~~~bpe~~~&~~~time [$\mu$s] \\
\hline
BV (7,3) & 2.063 & 0.21 & -- & -- \\
2a & 1.101 & 20.77 & 1.087 & 21.10 \\
4b & 0.909 & 29.03 & 0.890 & 27.43 \\
LM16 & 1.668 & 0.43 & 1.411 & 0.47 \\
LM32 & 1.320 & 0.55 & 1.228 & 0.69 \\
LM64 & 1.097 & 0.79 & 1.093 & 1.16 \\
\hline
\end{tabular}
\vspace{4mm}
\caption{Indochina-2004 dataset. Compression ratios (bpe) and access times per edge.
``LM$x$'' stands for LM-bitmap with $h = x$.
To the results of BV (7,3) the amount of 0.348\,bpe should be added, corresponding 
to extra data required to access the graph in random order.}
\label{table:indochina}
\end{table}

The larger experiment was run on four datasets (in both direct and transposed versions); 
the obtained results are presented in Fig.~\ref{fig:times} and exact numbers, for more 
careful examination, can be found in the appendix.
The LM-bitmap variant fares better in comparison with smaller blocks ($h$ up to 16), 
but then the LM-diff variant starts to win in compression, and the gap grows with 
growing $h$.
Unfortunately, decoding LM-diff blocks is also in most cases costlier, 
with 74\% maximum loss for Indochina-2004 direct, $h = 64$.
On average, its loss in speed to LM-bitmap is not, however, that big.

\begin{figure}[pt]
\centerline{
\includegraphics[width=0.49\textwidth,scale=0.8]{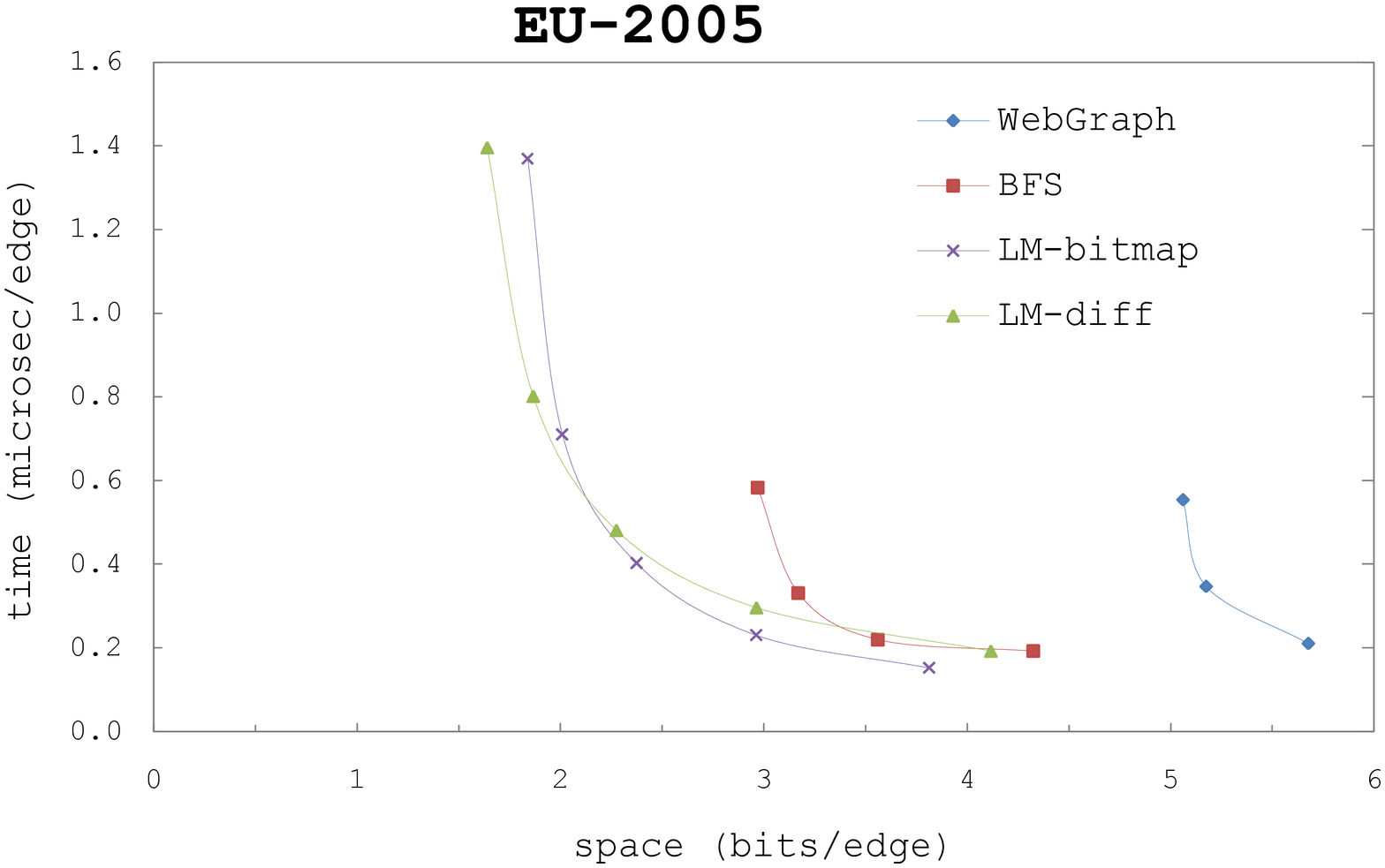}
\includegraphics[width=0.49\textwidth,scale=0.8]{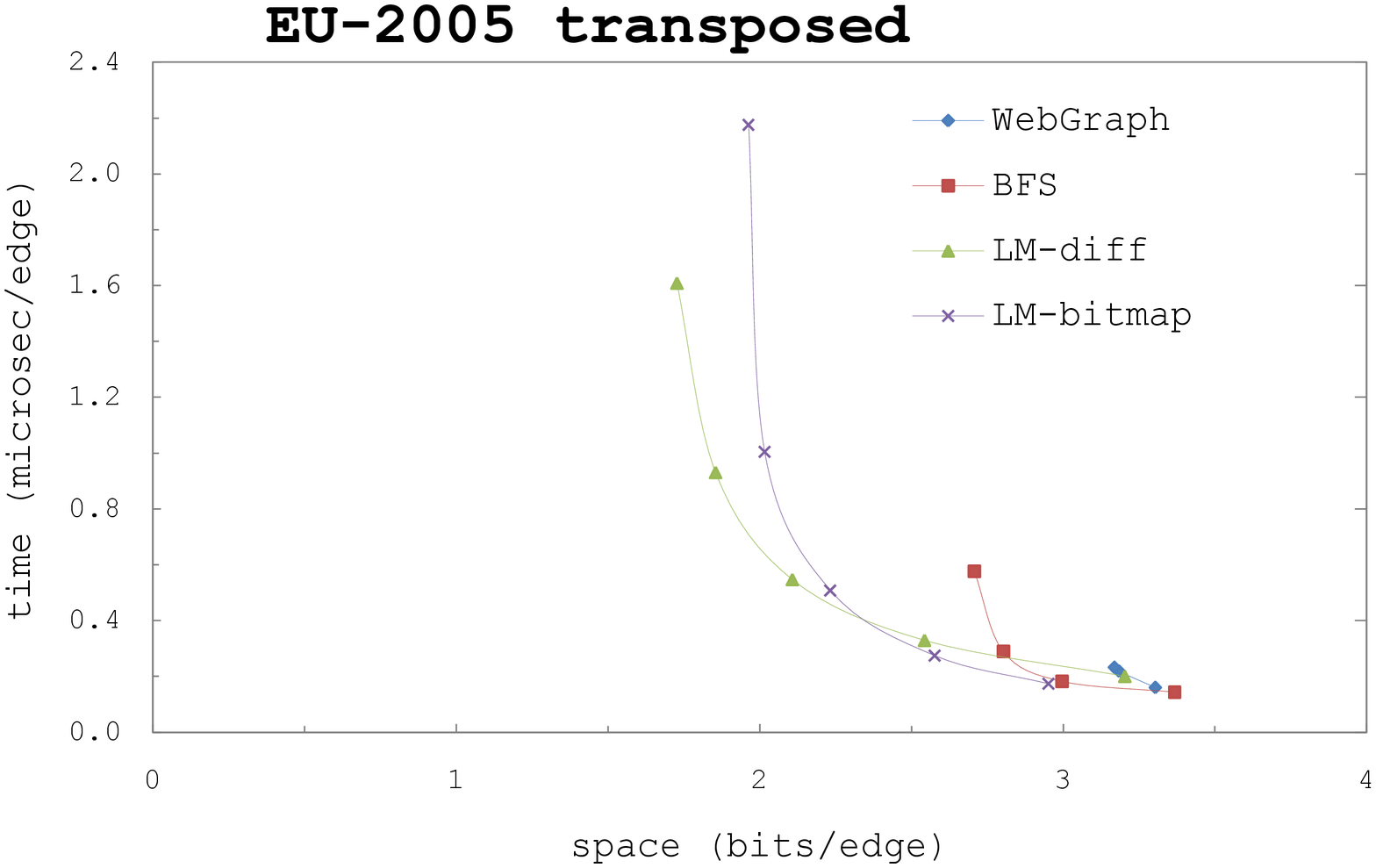}
}
\centerline{
\includegraphics[width=0.49\textwidth,scale=0.8]{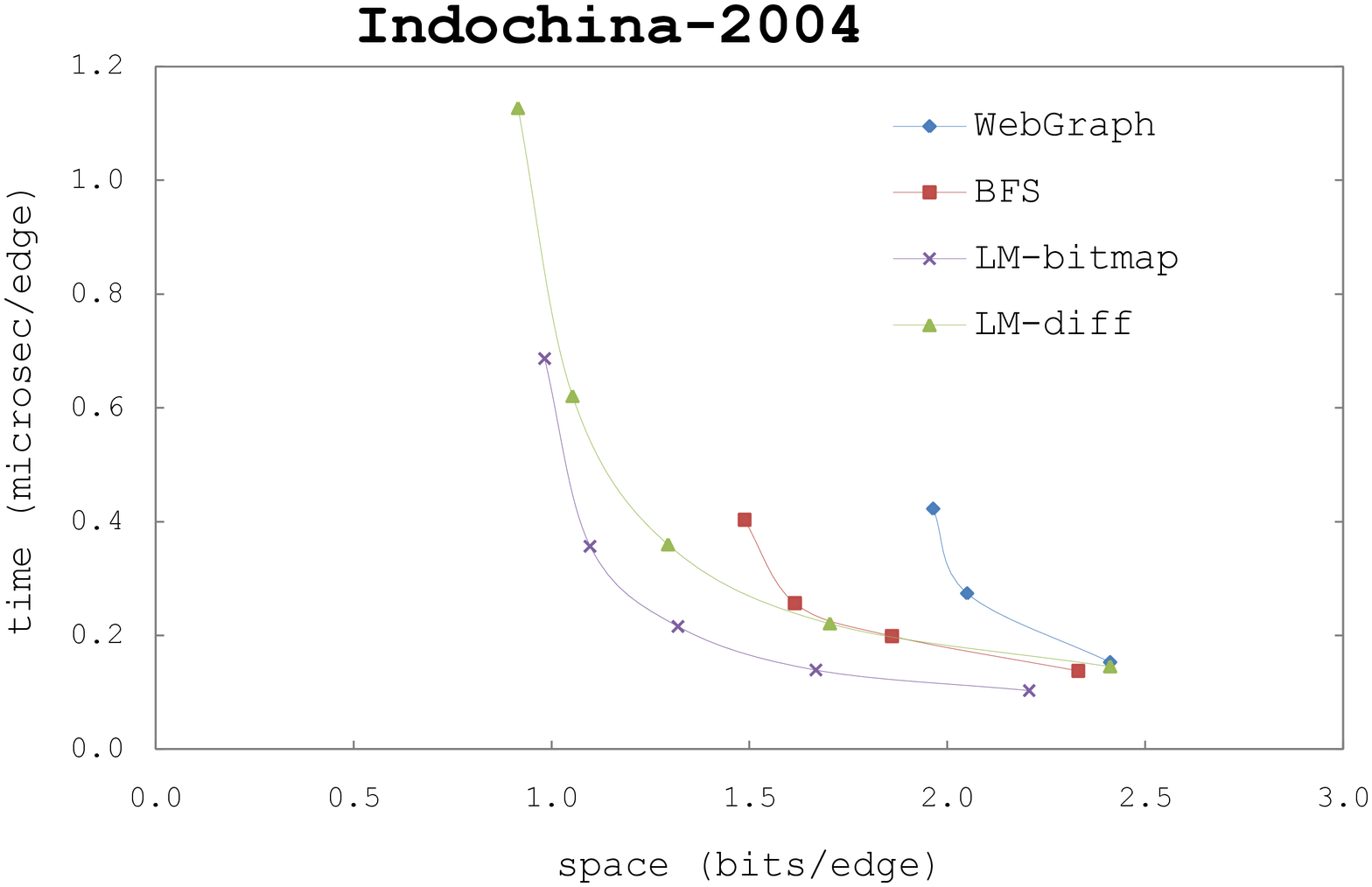}
\includegraphics[width=0.49\textwidth,scale=0.8]{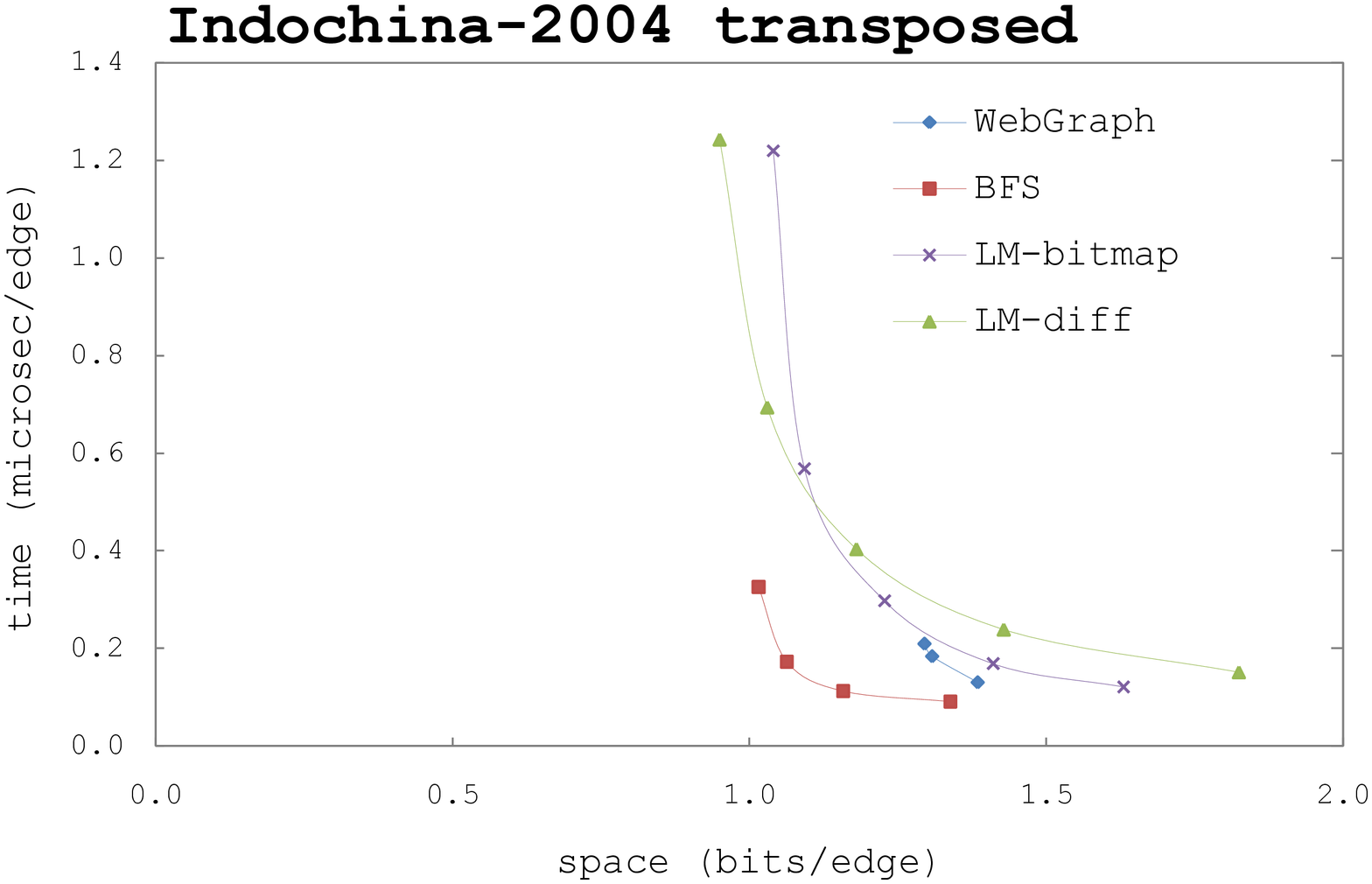}
}
\centerline{
\includegraphics[width=0.49\textwidth,scale=0.8]{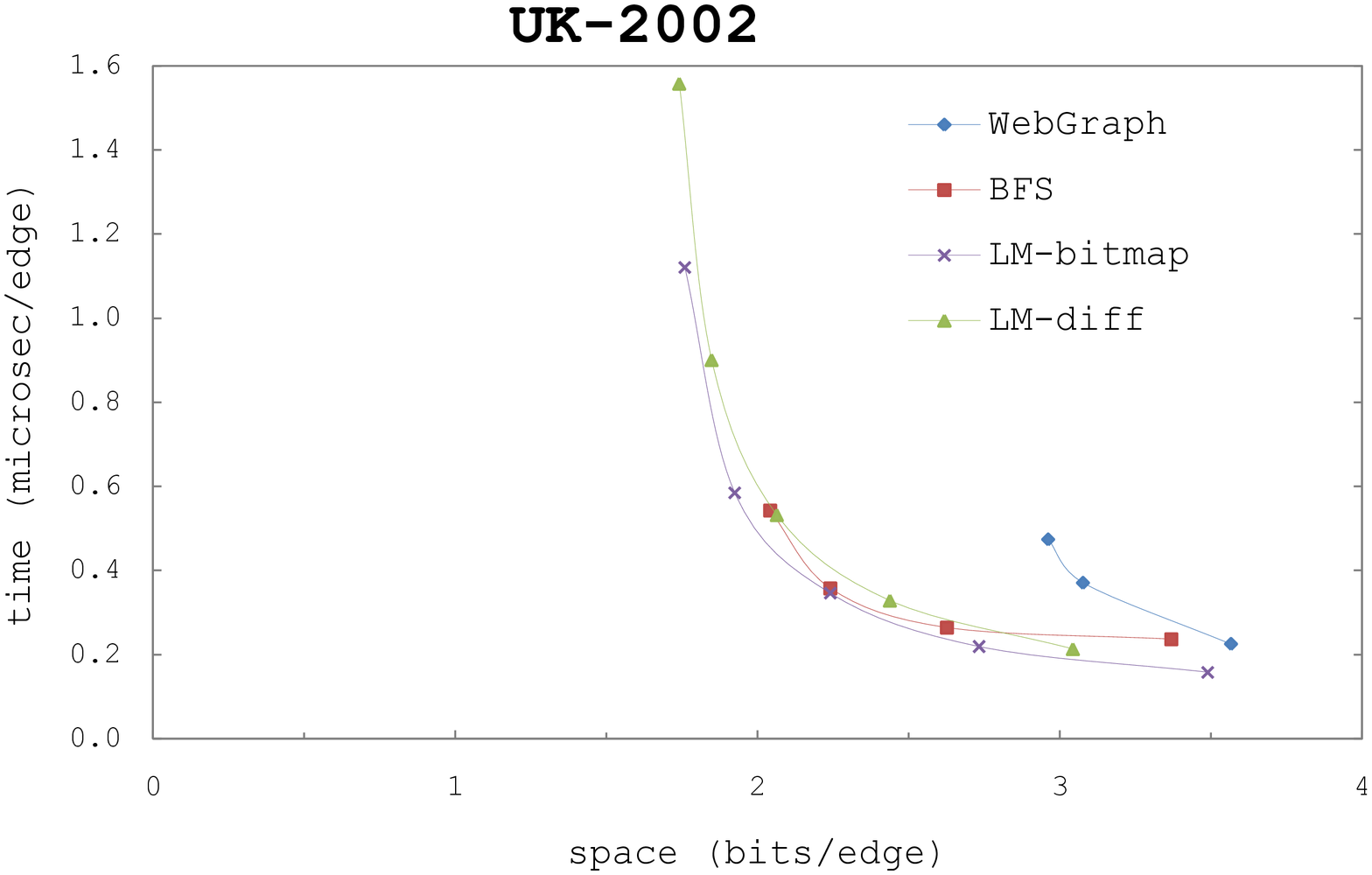}
\includegraphics[width=0.49\textwidth,scale=0.8]{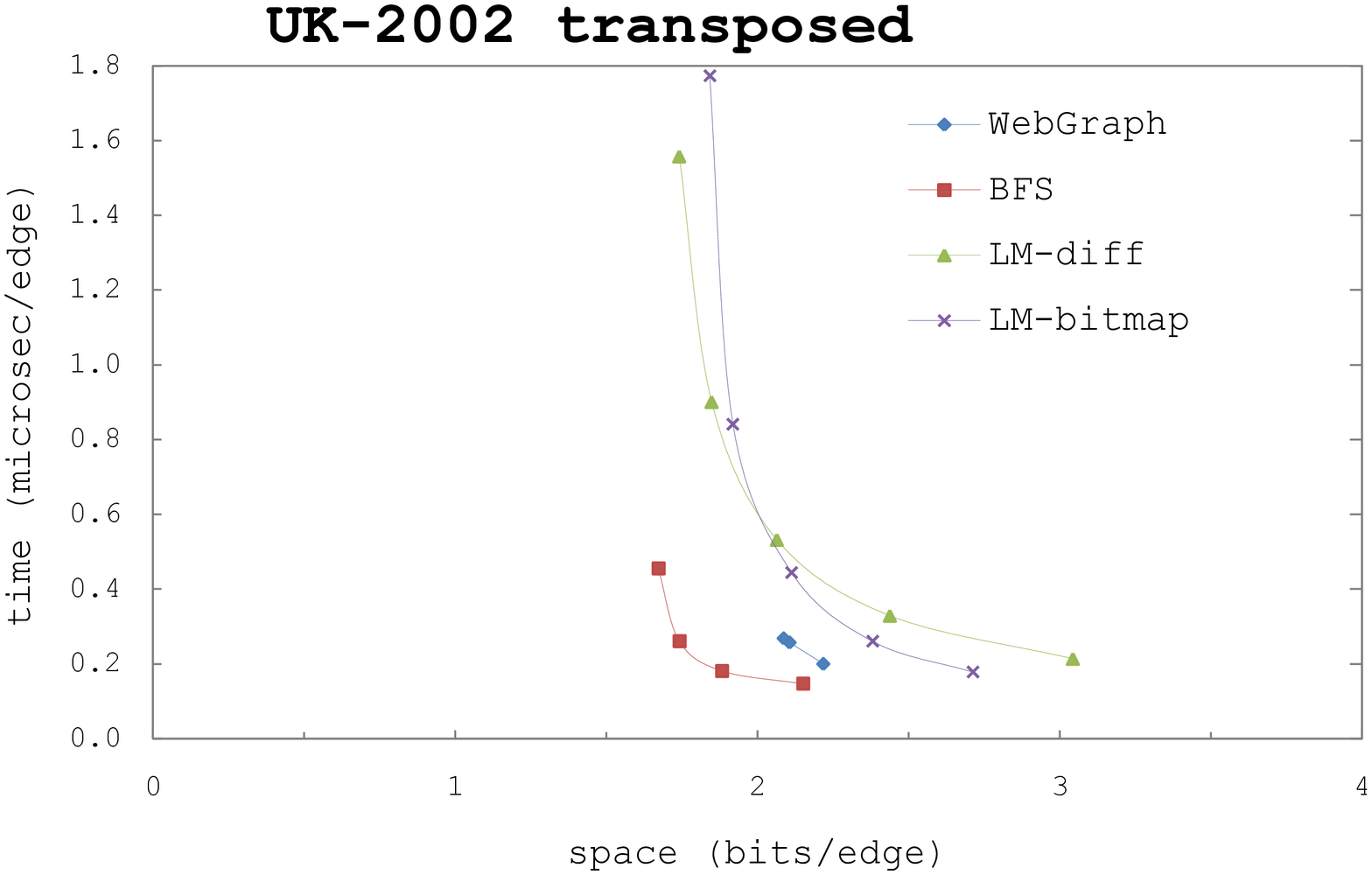}
}
\centerline{
\includegraphics[width=0.49\textwidth,scale=0.8]{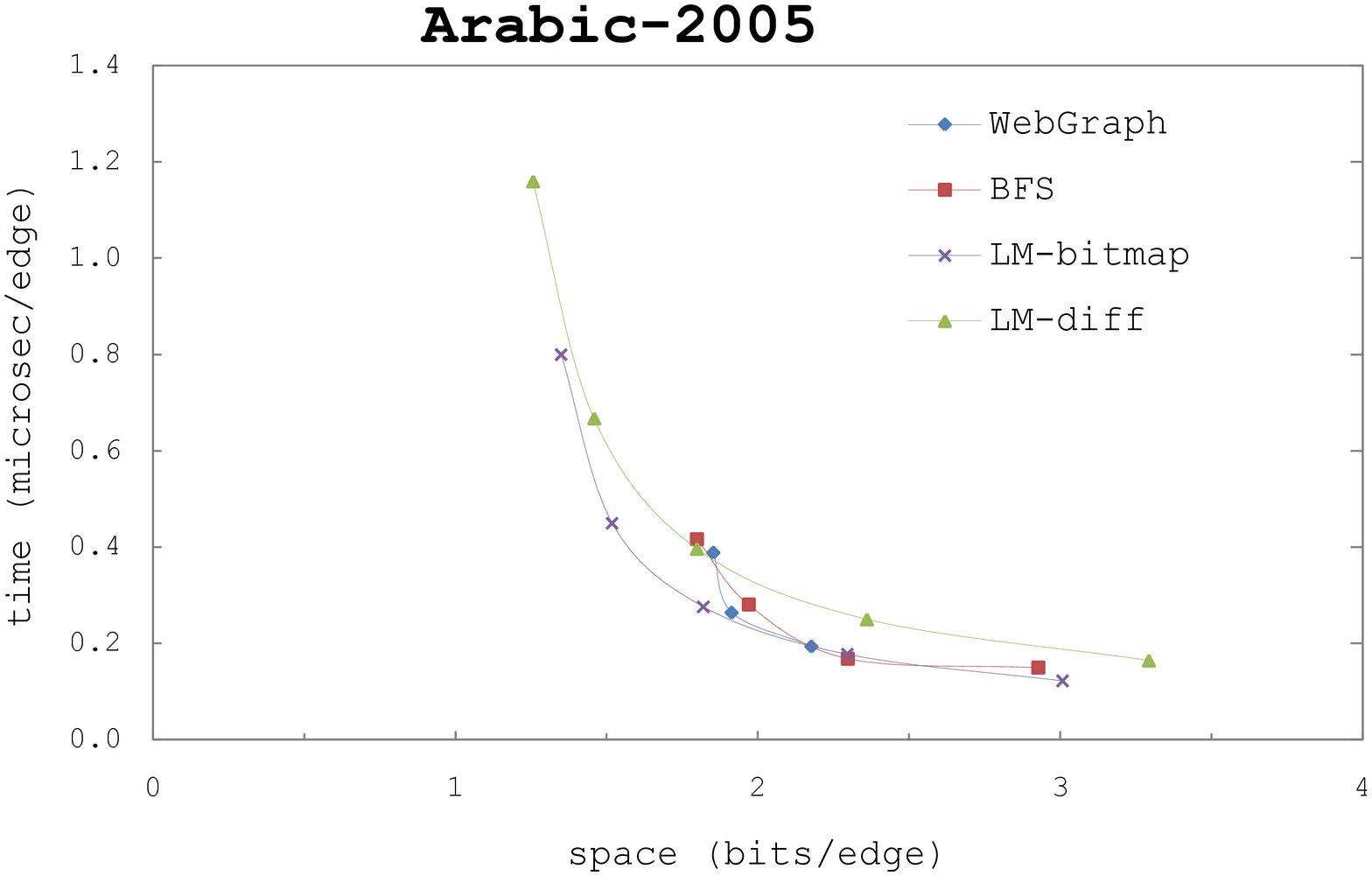}
\includegraphics[width=0.49\textwidth,scale=0.8]{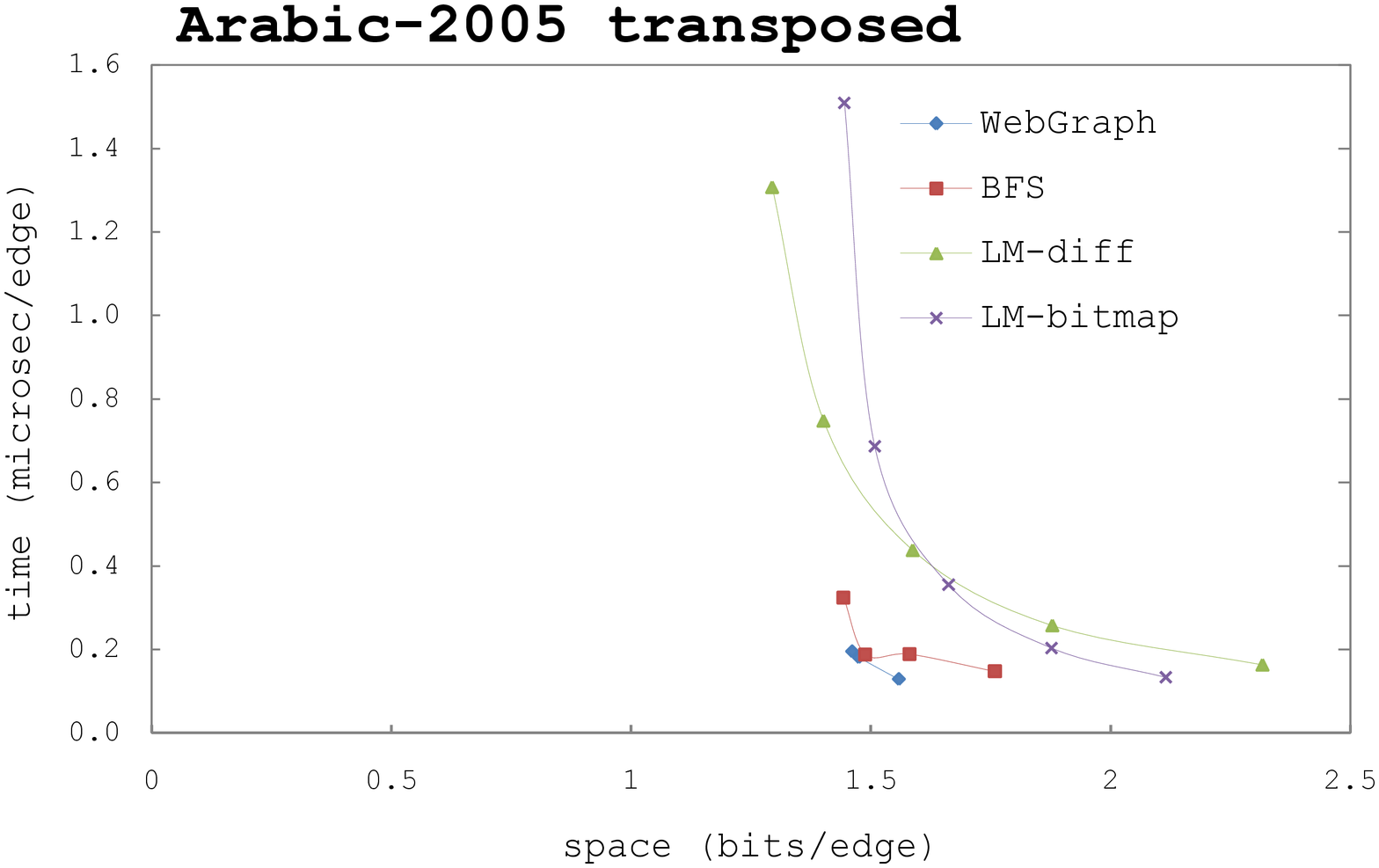}
}
\caption[Results]
{Compression ratios (bpe) and access times per edge}
\label{fig:times}
\end{figure}

%%%%%%%%%%%%%%%%%%%%%%%%%%%%%%%%%%%%%%%%%%%%%%%%%%%%%%%%%%%%%%%%%%%%%%%%%%%%%%
\subsection{Varying the block size in the algorithm based on similarity of successive lists}
%%%%%%%%%%%%%%%%%%%%%%%%%%%%%%%%%%%%%%%%%%%%%%%%%%%%%%%%%%%%%%%%%%%%%%%%%%%%%%

Obviously, the block size should seriously affect the overall space used by the 
structure and the access time. Larger blocks mean that the Deflate algorithm 
is more successful in finding longer matches and the overhead from encoding 
first lines in a block without any reference is smaller. On the other hand, 
more lines have to be usually decoded before extracting the queried adjacency list.

In this experiment we run the 2a algorithm (the same implementation in Java) with 
each block of residuals terminated (and later Deflate-compressed) after 
reaching BSIZE of 1024, 2048, 4096, 8192 and 16384 bytes, respectively. 
The test computer had an Intel Pentium4 HT 3.0\,GHz CPU, 1\,GB of RAM, 
and was running Microsoft Windows XP Home SP3 (32-bit).
The results (Table~\ref{table:blocks}) show that doubling the block size implies 
space reduction by about 10\% while the access time grows less than twice 
(in particular, using 8K blocks is only 2.0--2.5 times slower than using 2K blocks).
Still, as the block size gets larger (compare the last two rows in the table), 
the improvement in compression starts to drop while the slowdown grows.
For a reference, the access times of a practical Boldi--Vigna variant, BV (7,3), 
are 0.47\,$\mu$s and 0.42\,$\mu$s on the test machine.

\begin{table}
\centering
\begin{tabular}{rrrrr}
\hline
 & \multicolumn{2}{c}{EU-2005} &  \multicolumn{2}{c}{Indochina-2004}  \\
      ~~~&~~~bpe~~~&~~~time [$\mu$s]~~~&~~~bpe~~~&~~~time [$\mu$s] \\ 
\hline
1024 & 3.398 & 6.50 & 1.485 & 8.99 \\
2048 & 2.869 & 8.91 & 1.292 & 12.05 \\
4096 & 2.513 & 15.93 & 1.172 & 17.87 \\
8192 & 2.286 & 27.60 & 1.101 & 29.83 \\
16384 & 2.129 & 48.77 & 1.061 & 57.39 \\
\hline
\end{tabular}
\vspace{4mm}
\caption{Compression ratios and access times in function of the block size. 
2a variant used. Tests run on the non-transposed graphs.}
\label{table:blocks}
\end{table}

%%%%%%%%%%%%%%%%%%%%%%%%%%%%%%%%%%%%%%%%%%%%%%%%%%%%%%%%%%%%%%%%%%%%%%%%%%%%%%
\section{Conclusions}
%%%%%%%%%%%%%%%%%%%%%%%%%%%%%%%%%%%%%%%%%%%%%%%%%%%%%%%%%%%%%%%%%%%%%%%%%%%%%%

We presented two algorithms for Web graph compression, encoding blocks 
consisting of whole lines. 
All those algorithms achieve much better compression results than those 
presented in the literature, although two of them for the price of relatively slow 
access time. 
The more interesting algorithm, based on list merging, seems to be rather competitive 
to the algorithms known from the literature. 
%% (\cite{DBLP:conf/www/BoldiV04,CNtweb10,DBLP:conf/dcc/AnhM10}) 
%% but in experiments we could directly compare it only to the Boldi--Vigna algorithm.
Our approach lets achieve compression ratios not reported in the literature (LM-diff, 128), 
for one-directional queries, for moderate slow-down in list accesses
(the best tradeoff here, however seem to be the variants LM-diff and LM-bitmap 
for $h = 32$).

If even better compression ratios are welcome, then our SSL 4b variant can be 
considered, being more than an order of magnitude slower.
We point out that one extreme tradeoff in succinct in-memory data 
structures is when accessing the structure is only slightly faster than reading data 
from disk. The niche for such a solution is when the given Web crawl 
cannot fit in RAM memory using less tight compressed representation 
and the stronger compression is already enough.
The disk transfer rate is of relatively small imporantance here 
and what matters is the access time, which is about 10\,ms or more for commodity 
7200\,RPM hard disks.
Our algorithms spend significantly less time for extracting an average adjacency list, 
even if they are 1 or 2 orders of magnitude slower than the solutions from 
\cite{DBLP:conf/www/BoldiV04,CNtr08,DBLP:conf/birthday/ClaudeN10}.
Another challenge is to compete with SSD disks which are not much faster 
than conventional disks in reading or writing sequential data but their access 
times are two orders of magniture smaller. Here our LM variants are fast enough, though. 

%% Our future work will focus on improving the access times in both approaches; 
%% some possibilities lie in more aggressive reference list encoding via referring 
%% to several (cf. \cite{DBLP:conf/www/BoldiV04}) rather than a single previous list, 
%% using smaller independently compacted blocks with backend compression applied 
%% over many of them, and replacing Deflate with alternative compressors from LZ77 family, 
%% either stronger (e.g. LZMA, \url{http://www.7-zip.org/}), or even faster than Deflate 
%% in the decompression.

Our algorithm works locally.
In the future we are going to try to squeeze out some global redundancy 
while compressing the LM byproducts.
A natural candidate for such experiments is the RePair algorithm~\cite{LM00,CNtweb10}.
Other lines of research we are planning to follow are Web graph compression with 
bidirectional navigation and efficient compression of URLs.
As for bidirectional navigation, the very recent idea from Claude and Ladra~\cite{CL11} 
is a prospective approach, in combination with LM, but even summing up naively the sizes 
of the two structures we build now, for the direct and the transposed graph, 
gives quite interesting results (see~\cite{HNsnakdd11,CL11} for comparison).

\bibliographystyle{psc}
\bibliography{webgraph}

%%%%%%%%%%%%%%%%%%%%%%%%%%%%%%%%%%%%%%%%%%%%%%%%%%%%%%%%%%%%%%%%%%%%%%%%%%%%%%
\section*{Appendix}
%%%%%%%%%%%%%%%%%%%%%%%%%%%%%%%%%%%%%%%%%%%%%%%%%%%%%%%%%%%%%%%%%%%%%%%%%%%%%%

\begin{table}
\centering
\begin{tabular}{lrrrr}
\hline
        & \multicolumn{2}{c}{direct graph} & \multicolumn{2}{c}{transposed graph} \\
\cline{2-5}
        &  bpe~~~&~~~time [$\mu$s]~~~&~~~bpe~~~&~~~time [$\mu$s] \\
\hline
BV (7,3) & 5.679 & 0.211 & 3.304 & 0.160 \\
BFS, l4 & 4.325 & 0.192 & 3.367 & 0.144 \\
BFS, l8 & 3.561 & 0.219 & 2.996 & 0.183 \\
BFS, l16 & 3.169 & 0.330 & 2.803 & 0.289 \\
BFS, l32 & 2.969 & 0.583 & 2.708 & 0.576 \\
BFS, l1024 & 2.776 & 14.579 & 2.631 & 13.134 \\
LM-bitmap, 8 & 3.814 & 0.152 & 2.951 & 0.173 \\
LM-bitmap, 16 & 2.963 & 0.231 & 2.576 & 0.275 \\
LM-bitmap, 32 & 2.373 & 0.403 & 2.233 & 0.508 \\
LM-bitmap, 64 & 2.008 & 0.711 & 2.016 & 1.004 \\
LM-bitmap, 128 & 1.838 & 1.370 & 1.963 & 2.176 \\
LM-diff, 8 & 4.115 & 0.193 & 3.204 & 0.200 \\
LM-diff, 16 & 2.964 & 0.296 & 2.543 & 0.329 \\
LM-diff, 32 & 2.275 & 0.481 & 2.107 & 0.547 \\
LM-diff, 64 & 1.867 & 0.802 & 1.854 & 0.931 \\
LM-diff, 128 & 1.640 & 1.396 & 1.727 & 1.609 \\
\hline
\end{tabular}
\vspace{4mm}
\caption{EU-2005 dataset. Compression ratios (bpe) and access times per edge. 
All compressors are written in Java and were run with JRE 7.
The extra data required to access the graph in random order are included.
%% To the results of BV (7,3) the amount of 0.510\,bpe should be added, corresponding 
%% to extra data required to access the graph in random order.
}
\label{table:app_eu}
\end{table}

\begin{table}
\centering
\begin{tabular}{lrrrr}
\hline
        & \multicolumn{2}{c}{direct graph} & \multicolumn{2}{c}{transposed graph} \\
\cline{2-5}
        &  bpe~~~&~~~time [$\mu$s]~~~&~~~bpe~~~&~~~time [$\mu$s] \\
\hline
BV (7,3) & 2.411 & 0.153 & 1.384 & 0.130 \\
BFS, l4 & 2.331 & 0.137 & 1.339 & 0.091 \\
BFS, l8 & 1.860 & 0.199 & 1.158 & 0.112 \\
BFS, l16 & 1.615 & 0.257 & 1.063 & 0.173 \\
BFS, l32 & 1.488 & 0.403 & 1.016 & 0.326 \\
BFS, l1024 & 1.363 & 9.516 & 0.976 & 6.128 \\
LM-bitmap, 8 & 2.207 & 0.103 & 1.630 & 0.121 \\
LM-bitmap, 16 & 1.668 & 0.139 & 1.411 & 0.169 \\
LM-bitmap, 32 & 1.320 & 0.216 & 1.228 & 0.297 \\
LM-bitmap, 64 & 1.097 & 0.357 & 1.093 & 0.568 \\
LM-bitmap, 128 & 0.982 & 0.687 & 1.040 & 1.219 \\
LM-diff, 8 & 2.412 & 0.145 & 1.824 & 0.151 \\
LM-diff, 16 & 1.704 & 0.221 & 1.428 & 0.239 \\
LM-diff, 32 & 1.295 & 0.360 & 1.180 & 0.404 \\
LM-diff, 64 & 1.053 & 0.620 & 1.030 & 0.694 \\
LM-diff, 128 & 0.915 & 1.127 & 0.950 & 1.243 \\
\hline
\end{tabular}
\vspace{4mm}
\caption{Indochina-2004 dataset. Compression ratios (bpe) and access times per edge. 
All compressors are written in Java and were run with JRE 7.
The extra data required to access the graph in random order are included.
%% To the results of BV (7,3) the amount of 0.510\,bpe should be added, corresponding 
%% to extra data required to access the graph in random order.
}
\label{table:app_indochina}
\end{table}

\begin{table}
\centering
\begin{tabular}{lrrrr}
\hline
        & \multicolumn{2}{c}{direct graph} & \multicolumn{2}{c}{transposed graph} \\
\cline{2-5}
        &  bpe~~~&~~~time [$\mu$s]~~~&~~~bpe~~~&~~~time [$\mu$s] \\
\hline
BV (7, 3) & 3.567 & 0.225 & 2.218 & 0.200 \\
BFS, l4 & 3.369 & 0.236 & 2.152 & 0.147 \\
BFS, l8 & 2.627 & 0.264 & 1.883 & 0.181 \\
BFS, l16 & 2.242 & 0.357 & 1.742 & 0.260 \\
BFS, l32 & 2.042 & 0.542 & 1.673 & 0.455 \\
BFS, l1024 & 1.851 & 12.618 & 1.621 & 10.370 \\
LM-bitmap, 8 & 3.490 & 0.158 & 2.714 & 0.178 \\
LM-bitmap, 16 & 2.733 & 0.219 & 2.381 & 0.260 \\
LM-bitmap, 32 & 2.241 & 0.346 & 2.113 & 0.444 \\
LM-bitmap, 64 & 1.925 & 0.584 & 1.919 & 0.841 \\
LM-bitmap, 128 & 1.760 & 1.120 & 1.842 & 1.773 \\
LM-diff, 8 & 3.853 & 0.201 & 3.043 & 0.213 \\
LM-diff, 16 & 2.813 & 0.297 & 2.438 & 0.328 \\
LM-diff, 32 & 2.203 & 0.468 & 2.064 & 0.532 \\
LM-diff, 64 & 1.843 & 0.771 & 1.849 & 0.900 \\
LM-diff, 128 & 1.632 & 1.336 & 1.742 & 1.557 \\

\hline
\end{tabular}
\vspace{4mm}
\caption{UK-2002 dataset. Compression ratios (bpe) and access times per edge. 
All compressors are written in Java and were run with JRE 7.
The extra data required to access the graph in random order are included.
%% To the results of BV (7,3) the amount of 0.510\,bpe should be added, corresponding 
%% to extra data required to access the graph in random order.
}
\label{table:app_uk}
\end{table}

\begin{table}
\centering
\begin{tabular}{lrrrr}
\hline
        & \multicolumn{2}{c}{direct graph} & \multicolumn{2}{c}{transposed graph} \\
\cline{2-5}
        &  bpe~~~&~~~time [$\mu$s]~~~&~~~bpe~~~&~~~time [$\mu$s] \\
BV (7, 3) & 2.177 & 0.193 & 1.558 & 0.129 \\
BFS, l4 & 2.927 & 0.150 & 1.759 & 0.147 \\
BFS, l8 & 2.297 & 0.168 & 1.581 & 0.189 \\
BFS, l16 & 1.970 & 0.280 & 1.488 & 0.188 \\
BFS, l32 & 1.800 & 0.416 & 1.443 & 0.324 \\
BFS, l1024 & 1.631 & 12.327 & 1.408 & 8.692 \\
LM-bitmap, 8 & 3.008 & 0.122 & 2.116 & 0.133 \\ 
LM-bitmap, 16 & 2.295 & 0.177 & 1.877 & 0.203 \\
LM-bitmap, 32 & 1.820 & 0.276 & 1.662 & 0.355 \\ 
LM-bitmap, 64 & 1.518 & 0.449 & 1.508 & 0.687 \\
LM-bitmap, 128 & 1.350 & 0.799 & 1.445 & 1.509 \\ 
LM-diff, 8 & 3.293 & 0.164 & 2.317 & 0.163 \\
LM-diff, 16 & 2.361 & 0.250 & 1.879 & 0.258 \\
LM-diff, 32 & 1.798 & 0.396 & 1.587 & 0.438 \\ 
LM-diff, 64 & 1.459 & 0.667 & 1.401 & 0.748 \\ 
LM-diff, 128 & 1.256 & 1.159 & 1.294 & 1.307 \\
\hline
\end{tabular}
\vspace{4mm}
\caption{Arabic-2005 dataset. Compression ratios (bpe) and access times per edge. 
All compressors are written in Java and were run with JRE 7.
The extra data required to access the graph in random order are included.
%% To the results of BV (7,3) the amount of 0.510\,bpe should be added, corresponding 
%% to extra data required to access the graph in random order.
}
\label{table:app_arabic}
\end{table}

\end{document}